\documentstyle{l-aa}               
\input epsf.sty

\def\lapp{\mathbin{\raise2pt \hbox{$<$} \hskip-9pt \lower4pt \hbox{$\sim$}}}
\def\gapp{\mathbin{\raise2pt \hbox{$>$} \hskip-9pt \lower4pt \hbox{$\sim$}}}

\begin{document}
 
   \thesaurus{01         
              (02.16.2; 
               08.09.2; 
               08.14.2)} 

   \title{Phase Variability in the Optical Polarization of GRO~J1655--40}

  \author{ M. Gliozzi
           \inst{1} 
   \and    G. Bodo
           \inst{2}
   \and    G. Ghisellini
           \inst{3}
   \and    F. Scaltriti
           \inst{2,*}
   \and    E. Trussoni
           \inst{2}         
          }
  
  \thanks{Visiting Astronomer, Complejo Astronomico 
El Leoncito operated under agreement between the Consejo Nacional de
Investigaciones Cientificas y Tecnicas de la Republica Argentina and the
National Universities of La Plata, Cordoba and San Juan.}

   \offprints{G. Bodo}
 
   \institute
         {Dipartimento di Fisica Generale dell'Universit\`a, Via P. Giuria 1,
         I-10125 Torino, Italy
   \and  Osservatorio Astronomico di Torino, Strada
         dell'Osservatorio 20, I-10025 Pino Torinese (TO), Italy         
   \and  Osservatorio Astronomico di Milano, Sezione di Merate, Via Bianchi 46,
         I-22055 Merate (MI), Italy
                   }
 
\date{Submitted  ...; accepted  ...}
 
\maketitle

\markboth{M.\ Gliozzi et al.: Polarization of GRO~J1655--40}{}

\begin{abstract}
We present the results of the new optical polarimetric observations of the
superluminal source GRO~J1655--40, carried out in July 1997, with the
multichannel photopolarimeter of the Torino Observatory, using the 2.15-m
telescope of Complejo Astronomico El Leoncito (Argentina). The observed amount
of polarization  shows (mainly in the $V$ and $I$ bands) oscillations which are
consistent with the orbital period of the system. This gives information about
physical and geometrical properties of the binary system and confirms that the
optical continuum is intrinsically polarized. The origin of the polarized flux
is likely to be related to electron scattering of photons from the accretion
disk.


\end{abstract}

\maketitle

\section{Introduction}

\noindent
The microquasar  GRO J1655--40 (Nova Scorpii) has been one of the most studied
celestial objects in these last years both from the observational and the
theoretical point of view. It shows peculiar behaviour at all frequencies, from
radio to hard X-rays: in particular, thermal and non thermal emission and large
amplitude variability on short time scales. However the most important feature,
shared only with GRS 1915+105, is the ejection of radio emitting blobs with
superluminal motion. This phenomenology is interpreted as due to the activity
of an accretion disk surrounding a black hole of a few solar masses in a binary
system, with accreting material feeding the disk coming from a companion star
via Roche Lobe overflow (for a review of the main properties of Nova Scorpii
and GRS 1915+105 see Hjellming \& Rupen 1995 and Mirabel \& Rodriguez 1995). 

The observation of Bailyn et al. (1995b, $V$ band) can provide useful 
informations on the optical behaviour of the microquasar when it is active:

\noindent
{\it i)} the source is about 0.7-0.8 magnitude brighter in V with respect to the
quiescent phase ($16.3\leq V \leq 16.8$);

\noindent
{\it ii)} there is an exchange of the locations of the two minima during the
orbital period. In the bright state the primary minimum is at the orbital phase
$\phi=0.0$, and is due to the eclipse of the accretion disk by the secondary
star. 

\noindent 
{\it iii)} the phases of minima do not line up properly with the spectroscopic
phase (the difference in phase is as large as 0.05). This suggests  that during
the active status the disk is probably asymmetric. 

Further informations on the structure of GRO J1655--40 can be collected from
optical polarimetric observations. In July 1996, during a phase of high X-ray
and radio activity begun in May 1996, the optical flux from the source was
polarized by an amount of $2.5 \div  4$~\% in the $VRI$ bands, with the
polarization direction nearly parallel to the accretion disk plane, assumed
perpendicular to the jet axis (Scaltriti et al. 1997). 
    
New optical polarimetric observations of GRO J1655--40 were performed at the
2.15-m telescope of Complejo Astronomico El Leoncito (Argentina) with the five
channel photopolarimeter of Torino Observatory (Scaltriti et al. 1989) in the
period July 1-8, 1997, when the X-ray activity was high (flux $\sim 1$ Crab,
2-10 keV) but lower than in July 1996. With respect to the 1996 observations we
have followed in more detail the behaviour of the polarization parameters in
different bands with the orbital period. 
 
\begin{figure}[t]
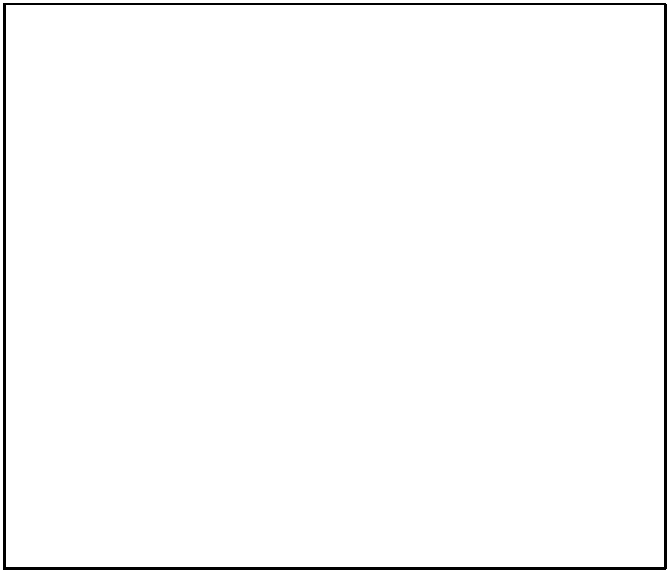

\picplace{7.5cm}
\caption{Plot of the observed fraction of polarization $P$ (left panel) and 
of the
position angle $\theta$ (right panel) as function of the
time (JD - 2450630.0) in the three bands $V$, $R$ and $I$ (from top to bottom).
The dotted lines are the average values.
The best fits to a sinusoidal curve are overplotted to the data of $P$
by fixing the oscillation amplitudes $P_{1,V} = 0.5$ \%, $P_{1,R} = 0.2$ \% and
$P_{1,I} = 0.8$ \%, respectively} 
\end{figure}
 
\section{Observations} 

\noindent
The observed magnitudes in the three bands $VRI$, the amount of polarization
$P$, the position angle $\theta$ and the orbital phase are reported
in Tab. 1 (the low flux in the $U$ and $B$ bands does not allow a useful
polarimetric analysis). A single determination of the polarization parameters
derives from rotations of 180$^\circ$ of the $\lambda/2$ plate over steps of
$22.5^{\circ}$. Each step lasts 10 s, and the whole cycle is repeated several
times, for a total run of $\approx 45$ minutes. Therefore each value of the
polarization parameters has been obtained over observational times 
corresponding to a fraction $\approx 0.015$ of the orbital period. 

Considering our photometric data, we can only infer that the source is in the
range of $15.1\leq V \leq 15.6$, about a magnitude brighter than the values
obtained during the active phase in March 1995 and roughly 0.7 magnitudes
fainter than the brightest value ever observed (Bailyn et al. 1995a). At JD
2450634 and 2450638 the sky conditions did not allow to derive reliable
photometry.

\begin{table*}
{\caption{(the errors are at 1$\sigma$ level)}}
\begin{tabular}{|c|c|c|c|c|c|}
\hline
FILTER   & MAG &  P(\%)  & $\theta$($^\circ$) & $\phi$ 
&  JD        \\
\hline
\hline
V        & 15.57$\pm$0.03 & 2.89$\pm$0.30 & 122.0$\pm$3.0 &  &    \\
R        & 14.92$\pm$0.03 & 3.69$\pm$0.34 & 126.0$\pm$2.7 & 0.562  & 2450631.6405 \\
I        & 14.03$\pm$0.03 & 3.06$\pm$0.43 & 124.8$\pm$4.0 &  &     \\
\hline
V        & 15.51$\pm$0.06 & 3.05$\pm$0.40 & 119.2$\pm$3.8 &  &     \\
R        & 14.88$\pm$0.04 & 2.98$\pm$0.24 & 129.9$\pm$2.3 & 0.937  & 2450632.6237  \\
I        & 13.96$\pm$0.07 & 2.46$\pm$0.52 & 131.6$\pm$6.0 &  &  \\
\hline
V        & 15.35$\pm$0.04 & 3.31$\pm$0.31 & 118.3$\pm$2.7 &  &     \\
R        & 14.81$\pm$0.03 & 3.50$\pm$0.38 & 121.6$\pm$3.1 & 0.312  &  2450633.6068  \\
I        & 13.78$\pm$0.05 & 3.83$\pm$0.66 & 121.4$\pm$4.9 &  &    \\
\hline    
V        & -- & 2.31$\pm$0.33 & 118.0$\pm$4.1 &  &  \\
R        & -- & 3.27$\pm$0.37 & 132.3$\pm$3.3 & 0.687 & 2450634.5899 \\
I        & -- & 2.32$\pm$0.39 & 124.8$\pm$4.8 &  &  \\
\hline
V        & 15.63$\pm$0.05 & 3.47$\pm$0.44 & 123.3$\pm$3.6 &   &     \\
R        & 15.24$\pm$0.07 & 3.12$\pm$0.32 & 126.0$\pm$3.0 &  0.062  &  2450635.5731  \\
I        & 14.29$\pm$0.05 & 2.85$\pm$0.62 & 131.5$\pm$6.1 &   &     \\
\hline 
V        & 15.50$\pm$0.04 & 3.09$\pm$0.30 & 119.4$\pm$2.8 &   &     \\
R        & 15.23$\pm$0.04 & 3.73$\pm$0.35 & 126.2$\pm$2.7 &  0.437  &  2450636.5562  \\
I        & 14.25$\pm$0.06 & 3.67$\pm$0.47 & 123.6$\pm$3.7 &      &     \\
\hline 
V        & 15.12$\pm$0.05 & 2.43$\pm$0.34 & 120.3$\pm$4.0 &    &     \\
R        & 14.87$\pm$0.05 & 2.87$\pm$0.24 & 131.7$\pm$2.4 &  0.812  &  2450637.5393  \\
I        & 13.85$\pm$0.05 & 1.99$\pm$0.35 &130.9$\pm$5.0 &     &     \\
\hline
V        & -- & 3.43$\pm$0.43 & 117.7$\pm$3.6 &    &     \\
R        & -- & 3.13$\pm$0.38 & 125.3$\pm$3.5 &  0.187  &  2450638.5224  \\
I        & -- & 3.53$\pm$0.70 & 123.5$\pm$5.6 &     &     \\
\hline
\end{tabular} 
\label{tab:table}
\end{table*}

\section{Results} 

\noindent
The temporal behaviour of $P$ and $\theta$ in the three bands is plotted in
Fig. 1. To verify that the observed fluctuations in $P$ are real we have
performed a $\chi^2$ test assuming for the observed polarization fraction a
sinusoidal fluctuation overlapped to a constant component: $P = \hat P + P_1 \,
{\rm sin} (2 \pi/T + \psi)$. In this expression $\hat P$ is the (weighted)
average value obtained in the three bands, $\hat P_V=2.98$ \%, $\hat P_R=3.18$
\%, $\hat P_I=2.84$ \%, while $P_1$, $T$ and $\psi$ are the amplitude, the
period (in days) and the phase of the oscillating component, respectively. We
remind that the orbital period of the system is $T \simeq 2.62$ days (van der
Hooft et al. 1998) and from our first observation we obtain $\psi=-0.401$. In
our tests we have assumed as parameters of the fit the amplitude and the period
of the oscillations. As only 8 data points are available we have carried out
the tests by fixing for each fit one of the two parameters (the reported errors
are at 68 \% confidence level for 7 degrees of freedom). 

\noindent
1) For $P_1=0$ (i.e. the hypothesis of constant $P$) the possibility
that the observed curve depends only on a statistical fluctuation is
$\lapp \, 16$ \% ($V$), $\lapp \, 32$ \% ($R$) and $\lapp \, 4$ \% ($I$). Fixing
$T$ at the orbital period we have found for the amplitude of the
oscillations $P_{1,V} = 0.5 \pm 0.2$ ($\chi^2=1.7$), $P_{1,R} = 0.2 \pm
0.2$ ($\chi^2 = 6.6$) and $P_{1,I} = 0.8 \pm 0.2$ ($\chi^2=3.6$). The
uncertainties are quite large, however we can see that at 99 \%
confidence we have as lower limits for the amplitudes $ 
P_{1,V} \gapp 0.1$ and $P_{1,I} \gapp 0.2$.

\noindent
2) To estimate the oscillating period we have fixed $P_1$ at the best values
deduced in the previous fits. We have found $T_V=2.6 \pm 0.1$ ($\chi^2=1.4$),
$T_R = 2.8 \pm 0.2$ ($\chi^2=5.0$) and $T_I = 2.7 \pm 0.1$ days ($\chi^2=2.1$).
Acceptable values are also obtained for the first upper 3 - 4 harmonics, while
no good fit is found  up to $T = 10$ days. These values are consistent,
within $ 1 \sigma$, with the orbital period. 

\noindent
3) The possible intrinsic differences of the temporal curves of $P$ in the
three bands have been checked,  assuming  as the best result the fit deduced in
the $I$ band. We have found that polarization data in $V$ and $R$ can be
consistent with the best fit in $I$ at $\lapp \, 20$ \% and $\lapp \, 4 \times
10^{-3}$ of probability, respectively. 

Concerning the temporal behaviour of the position angle, we found that the
observed variations of $\theta$ around its average values ($\hat \theta_V =
119.8^{\circ}$,  $\hat \theta_R = 127.4^{\circ}$,  $\hat \theta_I =
126.5^{\circ}$) can be ascribed to statistical fluctuations with probability
$\lapp \, 90$\% ($V$), $\lapp \, 13$\% ($R$) and $\lapp \, 75$\% ($I$). We have
further verified that the possible periodicity in the $R$ band has a period
of $1.6 \pm 0.1$  days. 

The results of these fits to the $P$ data have been overplotted to the observed
data in Fig. 1. From the quite accurate determination of the period we can
reasonably assume that the observed fluctuations in the $V$ and $I$ bands are
real and related to the orbital period of the system.  We remark that the low
values of the $\chi^2$ imply that our measurements are affected by systematic
errors higher than expected from statistical uncertainties alone. The estimate
of the amplitude of oscillations is quite poor, and the value in the $R$ band
could be assumed as an upper limit. It must be pointed out also that evidence
of polarization oscillations has been found in both the $I$ and $V$ bands, but
in $I$ the object is brighter by $\approx 1.5 \div 2$ magnitudes.

 \section{Discussion}

\noindent
Concerning the photometry, the most important implications of our data, when
compared with previous observations,  is that GRO J1655-40 is quite bright at
optical wavelengths in spite of its moderate X-ray activity. However the light
curve is not detailed enough for a reasonable comparison with previous
observations, when the source was in different activity phases. 

In order to gain informations about the physical and the  geometrical
properties of the binary system from the temporal fluctuations of the
polarization parameters, we have plotted in Fig. 2 the polarization fraction
vs the photometric phase. 

On the basis of the discussion in the previous Section, the observed fluctuations 
in the amount of
polarization seem to be 
related to the eclipses in the binary system.   In this case, the modulation of
$P$ can be due both to the occultation of the region from which the polarized
light originates and to  a variation of the non polarized flux due to the
eclipses in the system. One can then deduce that both the polarization region
and the covering medium must be extended, respectively because of the long fall
and rise time and of the smooth modulations in the amount of polarization. 
 
The most evident feature in the shapes of the polarization curves in all the
$VRI$ bands is the minimum of $P$ at $\phi \simeq 0.7 \div 0.8$, when the
accretion disk and the secondary star are roughly in quadrature. Note also the
lack of a similar minimum in the symmetric quadrature phase $\phi \sim 0.25$;
this may suggest that the system (and particularly the accretion disk)
is asymmetric. It is also worth noticing that the minimum of the polarized
amount, in the $VRI$ bands, occurs nearly at the same orbital phase of the
X-ray dips (Ueda et al. 1998, Kuulkers et al. 1998). Therefore it could be
related either to thickening of the disk rim, where the stream from the
companion hits the accretion disk (Parmar \& White 1988), or to the remnant
stream, which propagates inward above and below the disk after the impact with
the disk rim (Frank et al. 1987). 

\begin{figure}[t]
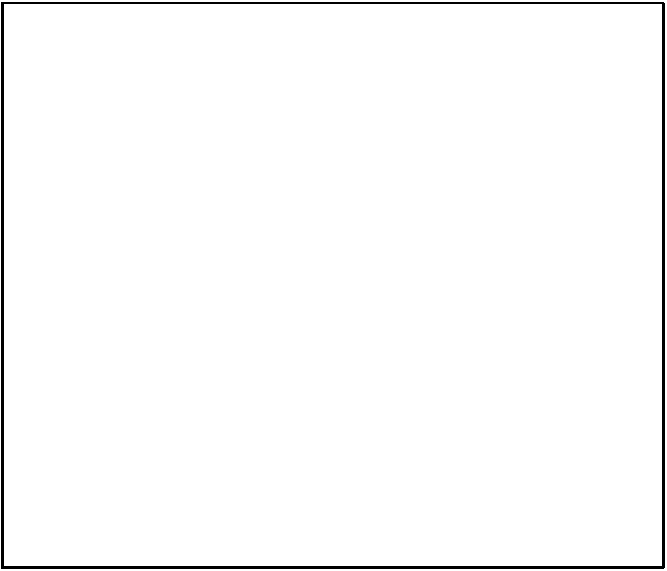

\picplace{7.5cm}
\caption{Plot of $P$ 
vs the photometric phase $\phi$ in the three bands $V$, $R$ and $I$ (from top
to bottom)
} 
\end{figure}

Another peculiar feature of $P(\phi)$ is that the maximum of polarized
fraction seems to lie in the phases $\approx 0 \div 0.5$, depending on the
band. Let us assume  that during the active state the accretion disk becomes
hotter than the companion star (as suggested in Bailyn et al. 1995b) and that
the region of emission of polarized photons is partially obscured in the
eclipse ($\phi \approx 0$). In the $I$ band the contribution of (unpolarized) photons from the
companion star to the total flux are prevailing, then we can expect the maximum
of polarization at $\phi \approx 0.5$ when the disk is fully visible and the
companion star partially obscured. This effect is further enhanced if the inner
hemisphere of the secondary (i.e. the one facing the compact object) is heated
up by the X-ray emission from the disk. 

In the $V$ band the behaviour is more complicated, as the diluting effect of
the flux from the companion should be reduced. Then at $\phi \approx 0$
the high energy photons from the disk (polarized and unpolarized) are partially
obscured, while all the disk is unveiled for $\phi \approx 0.5$. We could
justify the observed oscillations of $P$ in $V$ by assuming that the region of
emission of polarized flux reduces by increasing the photon energy, such
that when the disk is eclipsed only its unpolarized photons are obscured.
Furthermore at $\phi \approx 0.5$ the polarization fraction could be reduced by
the emission of the heated up hemisphere of the companion star. In such a case
the maximum of $P$ should be shifted to $\phi \approx 0$: this seems to be the
trend when comparing the $V$ and $I$ curved in Fig. 2. However our data do not
allow a strong support for this possibility. For intermediate wavelengths these
two different effects may compensate, leading to a much less evident
oscillations of the polarized flux, as observed in the $R$ band. 

In the previous paper (Scaltriti et al. 1997), we were not able to say whether
the origin of the polarized optical continuum was due to a nonthermal emission
process (synchrotron emission from the relativistic plasma inside the jet) or
to electron scattering. If the previous interpretations on the $P$
oscillations are reasonable, we can deduce now informations  on the physical
mechanism providing the polarized photons and on the geometry and location of
the region where polarized photons originate. The presence of smooth
eclipse-like modulations in the polarization curve implies that
the polarization region must be rather extended and lie, at least
partially, near the inner part of the accretion disk. The
same argument excludes, as a possible origin of the polarized optical
continuum, both the jet far away from the accretion disk, that is never
eclipsed, and the jet near the compact object, that cannot have a large spatial
extent. Therefore we conclude that the synchrotron emission from the jet cannot
be the physical mechanism originating the observed polarized radiation.
The most likely explanation is the electron scattering by plasma above the
accretion disk. It must be noticed that the presence of a highly ionized plasma
structure, similar to the one argued by our polarimetric observations, has been
recently confirmed by ASCA observations (Ueda et al. 1998). 

\section{Summary}

We have presented the results on new photopolarimetric observations of GRO
J1655--40, carried out in  July 1997, when the X-ray activity of the source was
still high but constantly decaying. These are the main conclusions: 

\begin{itemize}

\item With respect to July 1996 the source was brighter in optical 
($15.1\leq V \leq 15.6$ versus $V\sim16.2$), while the X-ray flux 
(2-10 keV) was $\approx 3$ times dimmer.

\item The observed fraction of polarization ($\approx 2.8 \div 3.2$ \%) shows
small amplitude oscillations in the $V$ and $I$ bands with a period consistent
with the orbital one. 

\item The curve $P(\phi)$ shows in all bands a minimum at $\phi \simeq 0.7\div
0.8$, interpreted as an asymmetric structure in the accretion disk.  The shape
of $P(\phi)$ in $VRI$ for $0 \lapp \phi \lapp 0.5$, is probably related to the
diluting effect of the unpolarized flux from both the star and the disk, and to
the partial eclipse of the region where the polarization photons originate. 

\item The smoothness in the polarimetric modulations suggests that the
polarization region is rather extended and close to the inner accretion disk.
Electron scattering from this zone seems to be the likely origin of the
polarized photons. 

\end{itemize}

\begin{acknowledgements} The authors wish to thank an anonymous referee for his
comments that greatly improved the presentation of these results. We also
acknowledge A. Spagna, S. Capetti and L. Feretti for discussions on the
statistical handling of temporal data. 
\end{acknowledgements}

\section{References}

\noindent{Bailyn C.D. et al., 1995a, Nature, 374, 701}

\noindent{Bailyn C.D. et al., 1995b, Nature, 378, 157}

\noindent{Frank J., King A.R.,  Lasota J.P., 1987, A\&A, 178, 137}

\noindent{Hjellming R.M., Rupen M.P., 1995, Nature, 375, 464}

\noindent{Kuulkers E. et al., 1998, ApJ, in press}

\noindent{Mirabel I.F., Rodriguez L.F., 1995, 17$^{\rm th}$ Texas Symp. on
relativistic astrophysics, New York Acad. of Sciences.}

\noindent{Parmar A.N.,  White N.E., 1988, Mem. Soc. Astr. It., 59, 147}

\noindent{Scaltriti F. et al., 1989, Mem. Soc. Astr. It., 60, 243}

\noindent{Scaltriti F. et al., 1997, A\&A, 325, L29}

\noindent{Ueda Y. et al., 1998, ApJ, 492, 782}

\noindent{van der Hooft F. et al., 1998, A\&A, 329, 538}

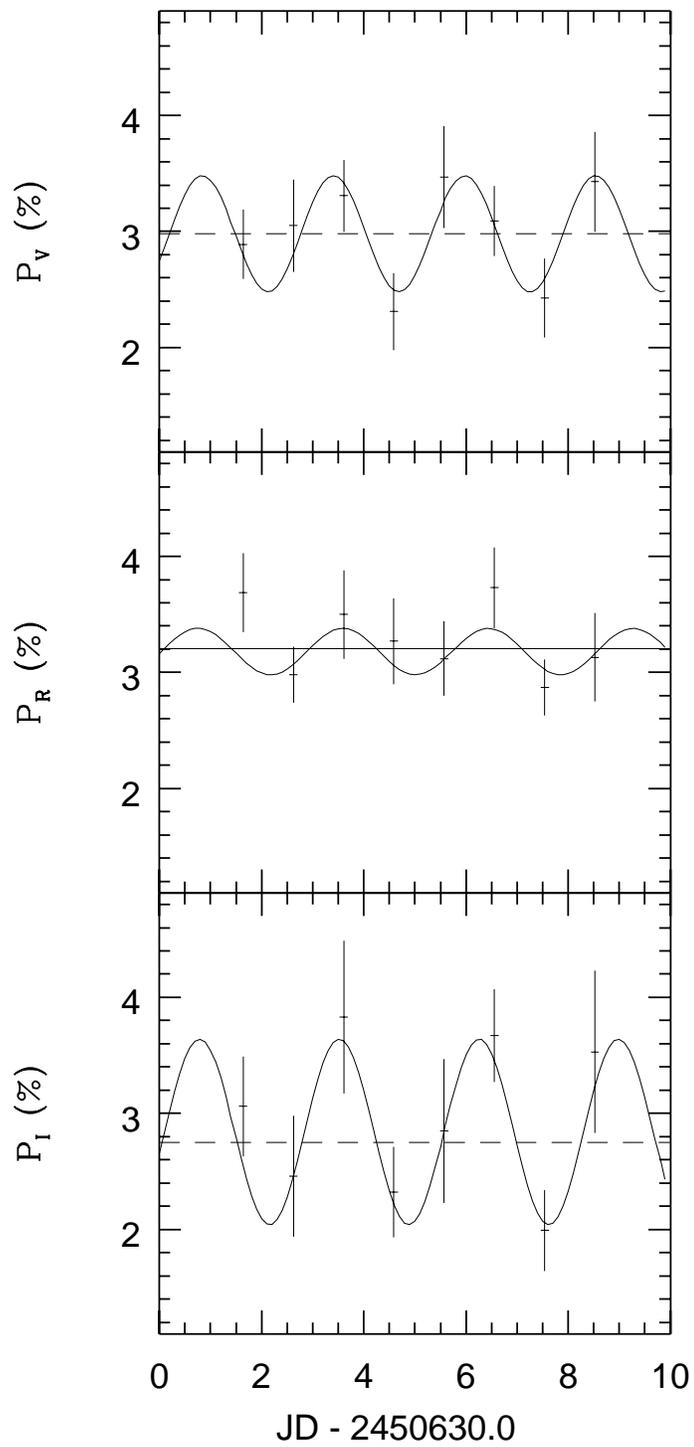
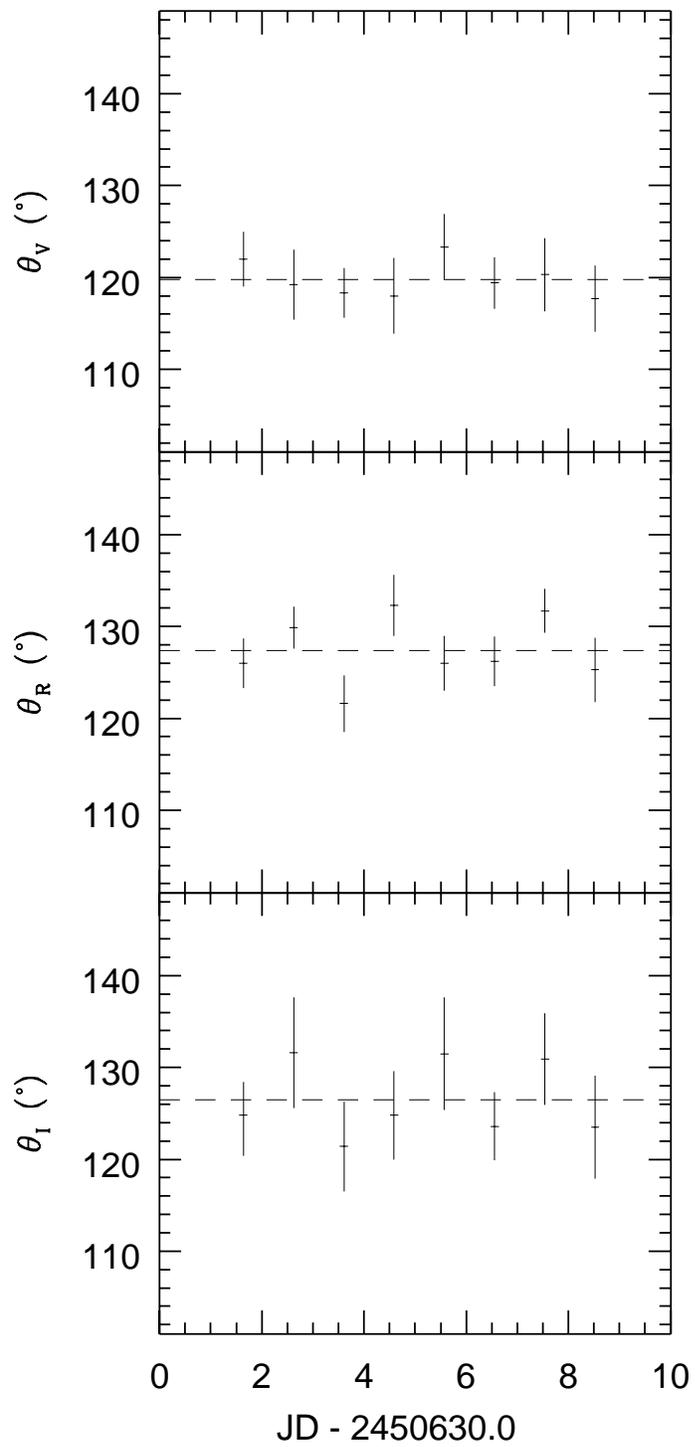
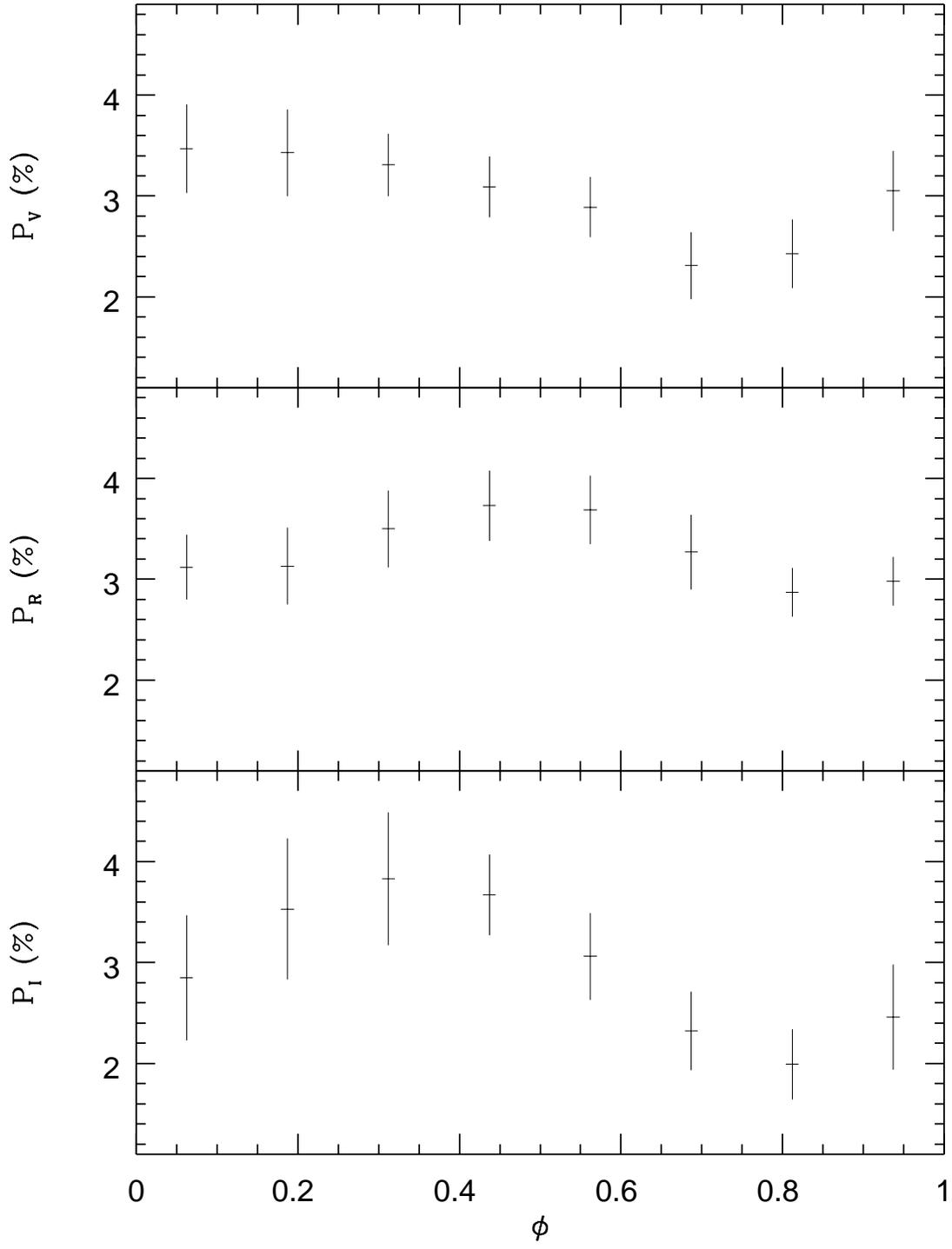
\end{document}